\documentclass[aps,prd,reprint,superscriptaddress,nofootinbib,floatfix,longbibliography]{revtex4-1}
\pdfoutput=1

\usepackage{amsfonts,amssymb,amsmath}
\usepackage{epsfig}
\usepackage{bm}
\usepackage{latexsym}
\usepackage{natbib}
\usepackage{url}
\usepackage{dcolumn}
\usepackage{color}
\usepackage{graphicx,epsfig}
\usepackage{hyperref}
\hypersetup{colorlinks   = true,
            urlcolor     = blue,
            citecolor    = blue,
            linkcolor    = blue,
            menucolor    = blue,
            anchorcolor  = blue,
            filecolor    = blue
}

\enlargethispage{\baselineskip}

\smallskipamount

\everymath{\displaystyle}

\begin{document}
\definecolor{myred}{RGB}{131,0,12}
\definecolor{dukeblue}{rgb}{0.0, 0.0, 0.61}
\title{A minimal Pati-Salam theory:\\ from cosmic defects to gravitational waves and colliders}

\author{Goran Senjanovi\'c}
\affiliation{
Arnold Sommerfeld Center, Ludwig-Maximilians University, Munich, Germany
}%
\affiliation{%
International Centre for Theoretical Physics, Trieste, Italy
}%
\author{Michael Zantedeschi}
\affiliation{INFN, Sezione di Pisa,
Largo Bruno Pontecorvo 3, I-56127 Pisa, Italy}
\email[]{ goran.senjanovic@physik.uni-muenchen.de\\ michael.zantedeschi@pi.infn.it}

\begin{abstract}
We discuss a minimal renormalizable Pati-Salam theory based on the $SU(4)_{\rm C}\,\times\,SU(2)_{\rm L}\,\times\,SU(2)_{\rm R}$ gauge group, with unification scale Higgs multiplets taken as $SU(2)_{\rm L}$ and  $SU(2)_{\rm R}$ doublets, 
which lead to neutrino Dirac picture. Although a number of scalar particles could be light, even lying at the LHC energies, the unification scale is hopelessly out of reach in any foreseeable future. Moreover,
 phase transition in the early Universe leads to the production of magnetic monopoles and domain walls, both incompatible with the standard cosmological model. A small explicit breaking of the discrete left-right symmetry allows the domain walls to decay, and in the process possibly sweep away the monopoles, analogously to the previously discussed case of $SU(5)$ grand unified theory. This leaves an important imprint of gravitational waves, within the reach of next generation searches, correlated with monopole detection and new light particles at collider energies. The theory has a dark matter candidate in the form of an inert scalar doublet, with a mass below TeV, which can further trigger electroweak baryogenesis.
\end{abstract}

\maketitle

\section{Introduction}

 Some fifty years ago Pati and Salam made a daring suggestion that quarks and leptons may be unified at high energies~\cite{Pati:1974yy}, thus paving the road to grand unification~\cite{Georgi:1974sy}. In their picture leptons are the fourth color, and the Standard Model (SM) interactions are unified within the $SU(4)_{\rm C}\,\times\,SU(2)_{\rm L}\,\times\,SU(2)_{\rm R} $ symmetry, augmented by the Left-Right (LR) symmetry which can be either parity $\mathcal{P}$ or charge conjugation $\mathcal{C}$. Strictly speaking the quark-lepton unification does not requite the LR symmetry~\cite{Pati:1973uk}, and indeed the LR symmetric aspect of the theory was  developed in subsequent works~\cite{Mohapatra:1974gc,Senjanovic:1975rk,Senjanovic:1978ev}.

 The $SU(4)_{\rm C}$ symmetry implies the existence of new gauge bosons, the leptoquarks 
\(X_{\rm PS}\), with mass given by scale of $SU(4)_{\rm C}$ symmetry breaking $M_{\rm PS}$, which couple quarks to leptons and mediate rare meson decays, such as say $K_L \to \bar e \mu$. Experimental limits on such processes require~\cite{Dolan:2020doe}
\begin{equation}
\label{eq:psconstrain}
    M_{\rm PS}\;\gtrsim\; 10^{5}\,\mathrm{GeV},
\end{equation}
roughly an order of magnitude above the LHC energies. 
However, the unification of color $SU(3)_{\rm C}$
and $U(1)_{\rm {B-L}}$ (baryon - lepton number) symmetry generically makes  $M_{\rm PS}$ orders of magnitude much larger, hopelessly out of direct experimental reach - leading to the million dollar question of how to test this theory. The absence of proton decay, the main signature of its cousin, the grand unified theory based on a single simple gauge group, would seem to offer no hope in foreseeable feature.

   Fortunately, the theory predicts the existence of magnetic monopoles~\cite{tHooft:1975psz}, in complete analogy~\cite{Polyakov:1974ek,tHooft:1974kcl} with grand unified theories - and their fate depends very much on the scale of partial unification. However, it also depends on the unknown cosmological evolution at high temperatures, so more is needed. The help comes from the existence of another type of topological defects, the domain walls, as a result of the spontaneous breaking of the discrete LR symmetry. These domain walls are a cosmological catastrophe within the standard cosmological model, independently of the scale of symmetry breaking - and so they must be eliminated one way or another. 
   
   Since the production of topological defects takes place through the high-temperature phase transition into an unbroken phase~\cite{Kibble:1976sj}, a simple way out would be to not restore the symmetry~\cite{Weinberg:1974hy, Mohapatra:1979qt,Mohapatra:1979vr,Mohapatra:1979bt}.
   As we will see, the symmetry non-restoration would seem to endanger the perturbativity of the theory, and thus we adhere to the conventional scenario with symmetries restored at temperature above the unification scale.
   
   Another solution of the domain wall problem is a tiny explicit breaking~\cite{Vilenkin:1984ib} of the discrete LR symmetry, even if suppressed by Planck scale effects~\cite{Rai:1992xw}. This case, albeit less appealing - one breaks the symmetry that one asks for - is more interesting since it forces domain walls to decay, and, in the process potentially sweep the monopoles away~\cite{Dvali:1997sa}. In the absence of cosmic inflation~\cite{Guth:1980zm} (and the minimal version of the theory does not inflate), this scenario becomes a necessity and thus we pursue it in this work.

In order to be able to do reliable computations we opt for the renormalisable version of the Pati-Salam (PS) theory. We choose the minimal Higgs sector based on the left and right weak doublets. This will imply Dirac neutrino masses, and in turn predict the absence of the violation of baryon (B) and Lepton (L) numbers.
Moreover, the simplicity of the Higgs sector leads to interesting relations of scalar particle masses and allows for the explicit form of the magnetic monopole solution, more of the reason to study this version of the minimal theory. The appeal of the Majorana picture and the resulting seesaw scenario does call for a dedicated study which we leave for the future. We comment on some potential implications in the outlook. 

We provide estimates indicating that the monopole problem is significantly alleviated, and possibly even resolved. Moreover, cosmological consistency of the thermal evolution of the Universe bears a clear signature: The production of a stochastic gravitational wave background with the frequency $f_{\rm GW}$ in the range $10\,{\rm Hz} \lesssim f_{\rm GW} \lesssim 10^3\, \rm Hz$ and a normalized energy density around $\Omega_{\rm GW} \equiv \rho_{\rm GW}/\rho_{\rm c}\simeq 10^{-7}$, where  $\rho_{\rm c}$ is the critical mass density of the Universe. 

Moreover, gravitational wave signal is intimately correlated with other phenomenological imprints. For example, a signal detected around $10^{3}\,{ \rm Hz}$ would point towards a monopole flux just below existing direct search bounds from MACRO~\cite{MACRO:2002jdv}. On the other hand, the lower frequency end would indicate the presence of a new scalar particle at present-day energies with Standard Model quantum numbers, in the obvious notation, $(3_{\rm C}, 2_{\rm L}, 7/6_{{\rm Y}/2})$. 

In this sense, this theory is self-contained:
besides providing correlations between magnetic monopoles, gravitational waves and present day (or near future) hadron colliders, it can further naturally address known cosmological conundrums, such as the origin of the dark matter and the genesis of its visible counterpart. 

Namely, somewhat surprisingly, a number of additional scalar states could also be light, and even lie at the present day or the near future collider energies, without spoiling the 
partial unification of gauge couplings. In other words, the usually assumed desert picture may not be true at all, and the resulting phenomenology could be relevant even for the LHC. These states comprise scalar analogs of a weak singlet down quark and a quark doublet, together with a number of the SM-like Higgs doublets. One of these doublets, the scalar analogue of the SM lepton doublet, is a natural Weakly Interacting Massive Particle, or WIMP as popularly called, (in the form of an inert weak doublet) dark matter candidate, and as such should lie at energies below TeV or so. Moreover, it (or some other scalar doublet) can help trigger first order electroweak phase transition and so lead to electroweak baryogenesis, which further leads to detectable gravitational waves, studied in the past. 

The rest of this paper is organized as follows. In the next Section we present the salient features of the theory and discuss the topological defects produced upon symmetry breaking. Sec.~\ref{whathappens} discusses the dynamics in the early Universe. In particular, it is shown that symmetry non-restoration poses a question regarding perturbativity of the theory, which leads to the cosmological catastrophes due to the production of both magnetic monopoles and domain walls. Their interaction is described in Sec.~\ref{sec:domainmonopolesweeping} where the phenomenon of monopole erasure is shown explicitly. Visuals of the dynamics can be found at the following \href{https://youtu.be/p7-EksvIrAs}{URL}. Sec.~\ref{cosmological consequences} deals with the cosmological consequences and signatures. Finally, Sec.~\ref{sec:concluion} is left for our conclusions and outlook.

\section{A minimal Pati-Salam model}
The PS theory is based on the gauge group 
\begin{equation}
\mathcal{G}_{\rm PS} \;=\; SU(4)_{\rm C}\,\times\,SU(2)_{\rm L}\,\times\,SU(2)_{\rm R} \times \mathcal{P},
\end{equation}
where $SU(4)_{\rm C}$ generalises the usual quark color symmetry to include leptons as the fourth color, and there is an additional discrete LR symmetry \footnote{We choose $ \mathcal{P}$ for definiteness. Our findings are analogous if charge conjugation $\mathcal{C}$ is realized instead.}. 

The minimal fermionic content requires three replicas of the following representations 
\begin{equation}
F_{\rm L} = (4_{\rm C},2_{\rm L},1_{\rm R}), \,\,\,
F_{\rm R} = (4_{\rm C},1_{\rm L},2_{\rm R}). \,\,
\end{equation}
These fields contain all the SM fermions, with the addition of the Right-Handed (RH) neutrino, predicting prophetically neutrino mass from the outset, decades before experimental confirmation. 

Regarding the Higgs sector, it is natural to ask for minimality, and we stick to it here. However, minimality does not suffice by itself, since we have an option of choosing either $SU(2)_{\rm L, R}$ doublets or triplets. The former lead to the Dirac neutrino picture, the latter to the Majorana one based on the seesaw mechanism. Over the years, the Majorana picture became dominant, but in the absence of any experimental hint of the neutrino Majorana feature, it is worthwhile to also consider the Dirac picture - after all, it was the original one~\cite{Mohapatra:1974gc,Senjanovic:1975rk}. 


\subsection{Higgs sector and symmetry breaking}

Besides the large scale Higgs fields, one needs the usual SM Higgs doublet, which finds its simplest embedding in the following complex bi-doublet $\Phi_1 = (1_{\rm C},2_{\rm L},2_{\rm R})$, where subscript 1 denotes the fact of it being $SU(4)_{\rm C}$ singlet. The bi-doublet must be complex, otherwise it would have the single Yukawa couplings and the single vacuum expectation value (vev), leading to the same masses of up and down quarks, and charged leptons and neutrinos. Unfortunately, this does not suffice since the SM Higgs doublet is the $SU(4)_{\rm C}$ singlet and thus cannot distinguish between leptons and quarks - leading to the wrong fermion mass predictions 
$m_u = m_\nu$, $m_d = m_e$. More is needed. 

The choice is simple: either we add higher-dimensional operator invoking $\langle H_{\rm R} \rangle$ - thus breaking the quark-lepton symmetry that leads to wrong mass relations at the $d=4$ tree level - or keep the renormalisable version by adding an additional complex bi-doublet field coupling to the fermions, $\Phi_{15} = (15_{\rm C},2_{\rm L},2_{\rm R})$, the adjoint representation of the 
$SU(4)_{\rm C}$ symmetry. It is easy to see that the joint presence of both $\Phi_{1}$ and $\Phi_{15}$
fields suffices to produce a realistic fermion mass spectrum, once the vevs of these fields turns on. The four independent Yukawa couplings guarantee the four independent quark and lepton mass matrices, allowing for usual CKM and PMNS mixing matrices. We shy away from the tedium of detailing the Yukawa structure, and refer instead the reader to for example~\cite{DiLuzio:2020xgc}.
In the rest of the paper, we  focus on the renormalisable version in order to be able to do explicit calculations of mass spectra and the forms of the topological defects, i.e. magnetic monopoles and domain walls. In the outlook we will comment on the non-renormalisable version.

In short, we have the following scalar multiplets:
\begin{equation}
\begin{split}
&H_{\rm L} = (4_{\rm C},2_{\rm L},1_{\rm R})\,, 
\quad H_{\rm R} = (4_{\rm C},1_{\rm L},2_{\rm R})\, ,\\
&\,\Phi_1 = (1_{\rm C},2_{\rm L},2_{\rm R})\,,
\quad\Phi_{15} = (15_{\rm C},2_{\rm L},2_{\rm R})\,,
\end{split}
\end{equation}
where $\Phi_1$ is a color singlet which, together with $\Phi_{15}$, generalizes the role of the Standard Model Higgs doublet. 

We start with the high-energy breaking of $\mathcal{G}_{\rm PS}$, hence the scalar potential takes the form
\begin{equation}
\label{eq:pot}
\begin{split}
    V_{H} = &- m^2 (H_{\rm L}^\dagger H_{\rm L} +H_{\rm R}^\dagger H_{\rm R} ) + \lambda \left[(H_{\rm L}^\dagger H_{\rm L})^2 + (H_{\rm R}^\dagger H_{\rm R})^2\right]  \\
    &+\lambda' \, {\rm Tr}(H_{\rm L}^\dagger H_{\rm L} H_{\rm L}^\dagger H_{\rm L}+H_{\rm R}^\dagger H_{\rm R} H_{\rm R}^\dagger H_{\rm R}) \\
    &+\lambda_{\rm LR} H_{\rm L}^\dagger H_{\rm L} H_{\rm R}^\dagger H_{\rm R}+\lambda'_{\rm LR}  {\rm Tr}(H_{\rm L}^\dagger H_{\rm L} H_{\rm R}^\dagger H_{\rm R}) \,,
\end{split}
\end{equation}
where the traces are over color indices. The scalar products $\overline{4}\,\times\,4 = 15 + 1$ leads to separate quartic couplings denoted by $\lambda,\lambda_{\rm LR}$ and $\lambda',\lambda'_{\rm LR}$ respectively. The sign of $\lambda_{\rm LR}'$ distinguishes orthogonal from parallel color embeddings of \(H_{\rm L}\) and \(H_{\rm R}\). Although it does not alter the vev structure, it turns out that $\lambda_{\rm LR}'$ determines residual symmetry in the domain wall, as discussed below.

It can be shown, as in the LR symmetric 
model~\cite{Senjanovic:1975rk,Senjanovic:1978ev}, that there is a portion of the parameter space with the vacuum configuration that breaks \(\mathcal{G}_{\rm PS}\) down to the Standard Model

\begin{equation}
\label{eq:vevextr}
     \langle H_{\rm R} \rangle \;=\;
    \begin{pmatrix}
        0 & 0 \\
        0 & 0 \\
        0 & 0 \\
        \dfrac{v_{\rm PS}}{\sqrt{2}} & 0
    \end{pmatrix},
    \qquad
    \langle H_{\rm L}\rangle \;=\; 0,
\end{equation}
yielding the gauge boson mass spectrum
\begin{equation}
\label{eq:gaugeboson}
    \begin{split}
        M_{\rm PS}^2\equiv M_{X_{\rm PS}}^2 =\frac{2\alpha_4}{2\alpha_2 + 3\alpha_4}
        M_{Z'}^2 = \frac{\alpha_4}{\alpha_2}M_{W_{\rm R}}^2= \alpha_4 \pi v_{\rm PS}^2 \,,
    \end{split}
\end{equation}
where $X_{\rm PS}$ denotes charge $\pm 2/3$ colored leptoquarks, $Z'$ the new neutral gauge boson, and $W_R$ the RH analog of the SM $W$ boson. 

Similarly, the scalar mass spectrum is given by
\begin{equation}
\label{eq:masses}
\begin{split}
    m_{\tilde{\nu}_R}^2 &= 2\,v_{\rm PS}^2\,\bigl(\lambda + \lambda'\bigr)\,,\\
    m_{\tilde{d}_R}^2 &= -\,v_{\rm PS}^2\,\lambda'\,,\\
    m_{\tilde{\ell}_L}^2 &= v_{\rm PS}^2\,\bigl(\lambda_{\rm LR} + \lambda_{\rm LR}' - 2\lambda - 2\lambda'\bigr)/2\,,\\
    m_{\tilde{q}_L}^2 &= v_{\rm PS}^2\,\bigl(\lambda_{\rm LR} - 2\lambda - 2\lambda'\bigr)/2\,,
\end{split}
\end{equation}
where we use a tilde to denote scalar fields, borrowing supersymmetric-like notation. Positivity of these squared masses is straightforward to achieve for suitable parameter choices. The imaginary part of \(\tilde{\nu}_R\) and the fields \(\tilde{u}_R,\tilde{e}_R\) become would-be Goldstone bosons, providing mass to the gauge bosons in the breaking \(\mathcal{G}_{\rm PS}\) down to the SM.

What about electroweak symmetry breaking? It proceeds in the usual manner through the Higgs doublets in $\Phi_1$ and $\Phi_{15}$. Although there are a number of such doublets, they generically end up being heavy due to the mixing with $H_{\rm R}$ - and in order to ensure the SM Higgs doublet, one must resort to the usual fine-tuning. In principle, though, more doublets could survive such tunings, and indeed, we will see below that the unification considerations do not restrict the number of light Higgs doublets as opposed to the situation in grand unified theories based on a single gauge group. 

\subsection{Pati-Salam scale}

As we said in the Introduction, the PS gauge lepqtouarks induce rare meson decays, resulting in the limit on the unification scale, $M_{\rm PS}\;\gtrsim\; 10^{5}\,\mathrm{GeV}$.
In contrast, typical renormalization group (RG) analyses push the PS scale much higher. By embedding color and \(B\!-\!L\) into \(SU(4)_C\) and enforcing left-right symmetry, one obtains
\begin{equation}
\label{eq:psunificationcondition}
\frac{1}{\alpha_1} 
    \;=\; \frac{3}{5}\,\frac{1}{\alpha_2} 
        \;+\; \frac{2}{5}\,\frac{1}{\alpha_3},
\end{equation}
where $\alpha_1$ coupling corresponds to the canonically normalized $U(1)$ generator $Y/2$. If one further assumes the so-called extended survival principles~\cite{Mohapatra:1982aq}, with the minimal fine tuning of just the SM Higgs doublet to be light (the popular desert picture), one finds
\begin{equation}
\label{eq:pssm}
    M_{\rm PS} \;=\; \exp\biggl\{\frac{\pi}{22}\,\Bigl(\frac{5}{\alpha_1}
        \;-\;\frac{3}{\alpha_2}\;-\;\frac{2}{\alpha_3}\Bigr)\biggr\}M_{\rm Z}\,,
\end{equation}
where the gauge couplings in parenthesis are evaluated at the $Z-$boson mass scale. 
Using the SM values \(\alpha_3^{-1}(M_Z)=8.4,\;\alpha_2^{-1}(M_Z)=29.6,\;\alpha_1^{-1}(M_Z)=59,\) the PS scale is $M_{\rm PS}\simeq\; 5\times 10^{13}\,\mathrm{GeV}$

However, things are not so simple. The extended survival principle may simply be wrong since new scalar states of the theory could lie much below the unification scale and impact $M_{\rm PS}$. Indeed, in some cases the particle spectrum is constrained by the requirement of a successful unification, leading to the violation of this principle~\cite{Bajc:2007zf,Preda:2022izo,Senjanovic:2024uzn,Preda:2024upl}. We are therefore required to relax this ad-hoc assumption. 

To quantify this, we perform the most general renormalization group analysis of such threshold effects. The scalar Higgs state are decomposed, under SM quantum numbers as
\begin{equation}
\begin{split}
    &H_{\rm L} = (3_{\rm C},2_{\rm L})_{\frac{1}{6}} +(1_{\rm C},2_{\rm L})_{\frac{1}{2}}\,,\\
    &H_{\rm R} = (\overline3_{\rm C},1_{\rm L})_{\frac{1}{3}} +(\overline{3}_{\rm C},1_{\rm L})_{-\frac{2}{3}} + (1_{\rm C},1_{\rm L})_{1}\,,\\
    &\Phi_1 = (1_{\rm C},2_{\rm L})_{\pm\frac{1}{2}}\,,\\
    &\Phi_{15} = (1_{\rm C},2_{\rm L})_{\pm\frac{1}{2}} + (\overline{3}_{\rm C},2_{\rm L})_{\pm\frac{1}{6}} + \\ 
    &\qquad \qquad\qquad+(\overline{3}_{\rm C},2_{\rm L})_{\pm\frac{7}{6}} + (\overline{8}_{\rm C},2_{\rm L})_{\pm\frac{1}{2}}\,.
    \end{split}
\end{equation}
For simplicity of presentation, we will consider the average mass scale of particle thresholds with the same quantum numbers. 
The unification condition \eqref{eq:psunificationcondition} leads to
\begin{equation}
\label{eq:psrgoneloop}
    \begin{split}
        &\frac{M_{\rm PS}}{M_Z} = \exp\left\{\frac{\pi}{21}\left( 
        \frac{5}{\alpha_1} - \frac{3}{\alpha_2} -\frac{2}{\alpha_3}\right)\right\}\times\\
        &\qquad\qquad\qquad \times\left[ \frac{(\overline{3}_{\rm C},2_{\rm L})_{\pm\frac{7}{6}}^{6}\, M_{\rm Z}}{(\overline{8}_{\rm C},2_{\rm L})_{\pm\frac{1}{2}}^4 \,(\overline{3}_{\rm C},2_{\rm L})_{\pm\frac{1}{6}}^3} \right]^{1/21}\,.
    \end{split}
\end{equation}
where we neglected the impact of $(1_{\rm C},1_{\rm L})_{1}$ and $(\overline{3}_{\rm C},1_{\rm L})_{-\frac{2}{3}}$ on the running since they are eaten by the heavy gauge bosons. Remarkably, other thresholds such as the Higgs doublets, and the analogs of the scalar down quark and lepton doublet drop out of the combination necessary for unification. In other words, these particle masses are arbitrary and for all that we know, they could be even accessible at present day or near future colliders. The usually assumed desert picture associated with unification need not be correct at all. 

Allowing then for the most general values of the thresholds, we arrive at the following range for the PS scale
    \begin{equation}
    \label{eq:mpslowerrunning}
  4 \times 10^{11}\,\rm GeV\, \lesssim M_{\rm PS} \lesssim 10^{18}\,\rm GeV\,. 
    \end{equation}
The lower limit is obtained
with the colored triplet state in the numerator of \eqref{eq:psrgoneloop} taken to be as light as roughly $10\,\rm TeV$, while the remaining states lie at the unification scale. 
On the other hand, the upper limit color octet and color triplet in the denominator to be light.

Notice that the $1-$loop RG analysis does not distinguish between gauge boson masses  and the unification scale. In order to be consistent - and to actually identify $M_{\rm PS}= M_{X_{\rm PS}}$ - a $2-$loop RG analysis is necessary. Its impact is discussed in Appendix~\ref{sec:2loop}.
We found that the $1-$loop unification is affected - and in particular, lowered - by less than $10\%$, 
leading to the lower bound on the PS scale. This correction is already included in \eqref{eq:mpslowerrunning}.

As it can be seen, the effect of the thresholds is dramatic: it allows $M_{\rm PS}$ to range over almost $7$ orders of magnitude. In other words, the theory fails completely to predict the unification scale, at first glance quite a blow to the PS program itself. However, it turns out that the cosmological considerations play a crucial role in this aspect. As we show in Sec.~\ref{subsec:erasure}, the validity of the standard cosmological model at the temperatures on the order of the unification scale narrows down the above range to a small interval $4\times 10^{11}\, {\rm GeV}\lesssim M_{\rm PS}\lesssim 10^{13}\,\rm GeV$.

\subsection{The role of left-right symmetry}

This modern day version of the original PS program has a twofold aim: first, to unify quarks and leptons and second, attribute the low energy maximal parity violation to the spontaneous breaking of the originally left-right symmetric theory. The former, in its minimal implementation, implies the $SU(3)_{\rm C}$ symmetry, while the latter requires doubling of the weak $SU(2)$ gauge group - and then augment it by the discrete $\mathbb{Z}_2$ exchange symmetries between these two $SU(2)$ symmetries. 

Here lies a catch, though. While gauge symmetries must be exact to start with in order for the theory to be renormalisable, the same is not true in case of global symmetries - a small breaking is certainly allowed, since the role of global symmetries is bookkeeping. So what do we mean here by the left-right symmetry? Or better, how to quantify the smallness of the breaking? 

It is actually not hard to answer this question. Spontaneous symmetry breaking keeps the memory of the original symmetry, due to its softness. For example, in our case the LH and RH quark mixing angles are equal to a good precision~\cite{Senjanovic:2014pva}, in spite of maximal parity breaking in beta decay. Similarly, neutrino Dirac matrices are almost hermitian, important for untangling the seesaw mechanism in the Majorana picture~\cite{Nemevsek:2012iq,Senjanovic:2016vxw} or having similar LH and RH lepton mixing angles~\cite{dartagnan}.  

In other words, when we speak of left-right symmetry, we simply assume that the explicit breaking is negligible compared to the spontaneous one. 
We will see, when we discuss the cosmology of the early universe, that these tiny explicit breaking terms  may indeed be necessary.

\subsubsection{Fermion masses and mixings}\label{subsubs:fermionmassandmixing}

Left-right symmetry leads to hermitian Yukawa couplings when LR symmetry is taken to be parity, and symmetric, when instead it is assumed to be charged conjugation. In the latter case, the quark mass matrices remain symmetric, and thus LH and RH quark mixing angles are predicted to be precisely the same. In the former case, the complex vevs of the bi-doublets spoil the potential hermiticity of mass matrices, but the memory of the LR symmetry remains and the mixing angles are the same for all practical purposes~\cite{Senjanovic:2014pva,Senjanovic:2015yea}. 

Therefore, in the quark sector, in order to maintain these predictions, all that is needed is that the explicit dimensionless parity breaking parameter $\epsilon$ be smaller than the lightest quark (up) Yukawa coupling, $Y_{\rm up} \simeq 10^{-5}$. If this breaking is due to quantum gravitational effects on the order $\epsilon \simeq (M_{\rm PS}/M_{\rm Pl})^2$, where $M_{\rm Pl}$ is the Planck scale, they would be negligible as long as $M_{\rm PS} \lesssim 10^{15}$ GeV - and we will see that the cosmological considerations will require $M_{\rm PS}$ much smaller, $M_{\rm PS} \lesssim 10^{13}$ GeV.

Since this theory leads to the Dirac picture, in the same manner, we would end up with similar LH and RH lepton mixing angles~\cite{dartagnan}, as long as 
$\epsilon \lesssim Y_{\nu} \simeq 10^{-9}$. In the case of quantum gravitational breaking effects, this translates into $M_{\rm PS} \lesssim 10^{13}$ GeV, in accord with cosmological constraints.

Of course, at this point these predictions of mixing angles are purely academic, due to large scale of LR symmetry breaking, far above experimental reach on the order $M_{\rm R} \simeq 10-100\,$TeV. Nonetheless, it is reassuring that the Planck scale suppressed breaking of LR symmetry, even if indeed present, does not affect any of the appealing features of the theory.

\subsubsection{Strong CP violation}

The underlying LR symmetry has interesting implications for the strong CP violation. At first glance, it would appear that it leads to a vanishing or tiny strong CP parameter $\bar \theta = \theta + \arg {\rm det}\, M_q$ - and indeed, the parity symmetry was invoked to claim naturally small strong CP violation~\cite{Beg:1978mt,Mohapatra:1978fy}. 

However, the situation is more subtle. Assuming no cancellation between  $\theta$ and $\arg {\rm det}\, M_q$, one is led to the requirement of vanishingly small phases of bi-doublet vevs~\cite{Maiezza:2014ala}. Similarly, the explicit parity breaking term must be negligible $\epsilon \lesssim \bar \theta \lesssim 10^{-10}$, which is satisfied for $M_{\rm PS} \lesssim 10^{13}$ GeV, as cosmological analysis would have it. It is pleasing that these rather different considerations lead to similar constraints. 

\subsection{Topological defects}
The theory is extremely rich, since it predicts the existence of both magnetic monopoles - due to the embedding of the $U(1)$ into the compact group - and domain walls - due to the breaking of LR discrete symmetry - as extended topological defects. 

\subsubsection{Magnetic Monopoles}
The high-energy phase transition in the PS model naturally produces `t~Hooft--Polyakov monopoles~\cite{tHooft:1974kcl,Polyakov:1974ek,tHooft:1975psz} of mass $\mathcal{O}(M_{\rm PS})$. These objects arise from the non-trivial topology of the vacuum manifold when $\mathcal{G}_{\rm PS}$ breaks down to the SM.

The explicit construction of these configuration is realized by Ref.~\cite{tHooft:1975psz} via the ansatz
\begin{equation}
\label{eq:monopolethooft}
    H_{\rm R}^m \;\propto\; \frac{v_{\rm PS}}{\sqrt{2}}
    \begin{pmatrix}
        0 & 0 \\
        0 & 0 \\
        e^{i\phi}\,\cos\!\tfrac{\theta}{2}\,\sin\!\tfrac{\theta}{2} 
        & \cos^2\!\tfrac{\theta}{2} \\[6pt]
        \sin^2\!\tfrac{\theta}{2}
        & e^{-i\phi}\,\cos\!\tfrac{\theta}{2}\,\sin\!\tfrac{\theta}{2}
    \end{pmatrix},
\end{equation}
which adopts the usual `t~Hooft--Polyakov asymptotic form in spherical coordinates $(r,\theta,\phi)$, thereby forcing a monopole at the origin~\cite{tHooft:1975psz}. Here, the proportionality sign is due to the fact that we omit the detailed radial profile. This expression is simply a combined $SU(2)_{\rm R}$ and $SU(4)_{\rm C}$ rotation on the vacuum state in~\eqref{eq:vevextr}, just as the original 
't Hooft-Polyakov monopole~\cite{tHooft:1974kcl,Polyakov:1974ek} is the $SO(3)$ rotation on the analogous vacuum state. 

A standard energy argument shows that, to cancel infinities at large $r$, both the $SU(4)_{\rm C}$ leptoquark gauge fields $X_{\rm PS}$ and the $SU(2)_{\rm R}$ gauge fields $W_{\rm R}$ must align appropriately with $H_{\rm R}^m$. This feature is essential to our subsequent discussion, as the presence of both colored and right-handed gauge bosons in the monopole core will enable the interaction with domain walls that can sweep the monopoles away.

\subsubsection{Domain Walls}\label{subsubsec:dws}
A second important defect emerges from the spontaneous breaking of left-right (LR) discrete symmetry, which produces domain walls (for a discussion of such domain walls in the LR symmetric model see~\cite{Ringe:2024ktt}). This is true as long as the explicit breaking terms are negligible, as we discuss above.
We label the left- and right-handed scalar multiplets as $H_{{\rm L}\,a i}$ and $H_{{\rm R}\,a i}$ (with $a=1,\dots,4$ for $SU(4)_{\rm C}$ and $i=1,2$ for $SU(2)_{\rm LR}$). A typical planar wall at $x=0$ can be approximated by
\begin{equation}
\label{eq:dwansatz}
\begin{split}
    &H_{{\rm L}\, ai} = -\frac{v_{\rm PS}}{2\sqrt{2}}\, \left(\tanh{[ m_{\rm DW}\,  x]} -1\right)\delta_{ab}\delta_{ij} \,,\\ &H_{{\rm R}\, 41} = \frac{v_{\rm PS}}{2\sqrt{2}}\, \left(\tanh{[ m_{\rm DW}\,  x]} +1\right)\,,  
    \end{split}
\end{equation}
where $m_{\rm DW}$ is the domain-wall inverse width.

The order parameter 
\begin{equation}
H_{\rm L}^\dagger H_{\rm L} \;-\;H_{\rm R}^\dagger H_{\rm R}
\end{equation}
interpolates in the usual manner from $v_{\rm PS}^2/2$ at $x \to -\infty$ to $-v_{\rm PS}^2/2$ at $x \to +\infty$, vanishing at the wall’s center. Inside the wall, both $H_{\rm L}$ and $H_{\rm R}$ become nonzero, so the local gauge group is at most $SU(3)_{\rm C} \times U(1)_{\rm {em}}$ (more on it below).

We can use ansatz \eqref{eq:dwansatz} to estimate the tension of the domain wall $\sigma$. Notice that the profile in~\eqref{eq:dwansatz} here is an approximation of the actual numerical solution. We verified that it properly approximates the actual numerical solution - at least for couplings of $\mathcal{O}(1)$ we were able to implement. To obtain the domain-wall tension $\sigma$, it suffices to integrate the energy density of the configuration~\eqref{eq:dwansatz} along the $x$ direction. For simplicity, we first comment on the case of positive $\lambda_{\rm LR}'$, for which $H_{\rm LR}$ are orthogonal in $SU(4)_{\rm C}-$space along the domain wall support. As we will see, this will not alter our results.

Direct integration of the energy density leads to
\begin{equation}
\begin{split}
    &\sigma_{\rm Wall} = \int {\rm d}x\, \mathcal{E}_{
    \Phi_{\rm DW}} =    
    \frac{m_{\rm DW}\,v_{\rm PS}^2}{3} + \frac{v_{\rm PS}^4\left[\lambda_{\rm LR} + 2\left(\lambda + \lambda'\right) \right]}{48 \,m_{\rm DW}}
    \end{split}
\end{equation}
where $m_{\rm DW}^{-1}$ is the width of the wall in~\eqref{eq:dwansatz}. Extremizing we get
\begin{equation}
    m_{\rm DW} = \frac{m_{\tilde\nu _R}}{4}\sqrt{1 + \frac{\lambda_{\rm LR}}{2(\lambda + \lambda')}} > \frac{\sqrt{2}}{4}m_{\tilde \nu_R}\,,
\end{equation}
where in the last equality positivity of the mass spectrum~\eqref{eq:masses} has been used to replace $\lambda_{\rm LR}$. The above expression is a straightforward generalization of the usual domain-wall width in the case of a single real scalar field - there it is the scalar particle mass, while here it is just a lower limit. 

Finally, we arrive at the result
\begin{equation}
\begin{split}
    \label{eq:sigmalower}
    \sigma_{\rm Wall} &= \frac{1}{6}v_{\rm PS}^3\sqrt{2(\lambda + \lambda') + \lambda_{\rm LR}}\\
    &=\frac{M_{\rm PS}^2\,m_{\tilde \nu_R}}{6\pi\, \alpha_{\rm PS}}\sqrt{1 + \frac{\lambda_{\rm LR}}{2(\lambda + \lambda')}}\,,
    \end{split}
\end{equation}
where we remind the reader $m_{\tilde \nu_R} = v_{\rm PS}\sqrt{2\left(\lambda + \lambda'\right)}$.

Using again positivity of the mass spectrum~\eqref{eq:masses}, combined with the fact that $M_{\rm PS}^2 = \pi\, \alpha_{4}\, v_{\rm PS}^2$ is the gauge mass of the PS gauge bosons $X_{\rm PS}$, we arrive at
\begin{equation}
    \label{eq:sigmalower}
    \sigma_{\rm Wall}  \geq \frac{M_{\rm PS}^2\, m_{\tilde\nu_R}}{3 \sqrt{2}\pi\, \alpha_4} \simeq 
    3 \, M_{\rm PS}^2\,m_{\tilde \nu_R}\,,
\end{equation}
where in the last inequality we used $\alpha_{4}\simeq 37^{-1}$ corresponding to the typical value of $\alpha_{4}$ at the unification scale $M_{\rm PS}\simeq 10^{13} \,\rm GeV$. 

If $\lambda_{\rm LR}'<0$, $H_{\rm L,R}$ are parallel in $SU(4)_{\rm C}$-space along the domain wall profile. In this case the tension is obtained by simply replacing $\lambda_{\rm LR}\rightarrow \lambda_{\rm LR} + \lambda_{\rm LR}'$. Notice that positivity of scalar spectrum \eqref{eq:masses} still leads to inequality \eqref{eq:sigmalower}.  

What about the physical implication of the sign of $\lambda'_{\rm LR}$? The relative orientation between $H_{\rm L,R}$ in $SU(4)_{\rm C}$ space determines the residual symmetry along the domain wall support. In particular, it informs us wether $SU(3)_{\rm C}$ is broken or not. For $\lambda'_{\rm LR}>0$, $SU(2)_{\rm C}\times U(1) \times U(1)$ gauge symmetry localizes on the wall. In the opposite case of $\lambda'_{\rm LR}< 0$, $SU(3)_{\rm C}\times U(1)_{\rm em}$ remains unbroken. This obviously has consequences regarding the behavior of these extended objects when evolving in the primordial plasma. 

Naively, one might think that the wall tension and $m_{\tilde\nu_R}$ can be arbitrarily small. While phenomenologically true, viability of the  standard cosmological scenario imposes a lower bound on the mass of the singlet, therefore forcing the tension to be near the GUT scale value. The reason for this is that for too small singlet mass, the radiative corrections to the potential~\cite{Coleman:1973jx} stemming from the gauge sector, can not only make the unbroken phase a local minimum, but the global one\footnote{We refer the interested reader to e.g.,~\cite{Daniel:1980xn} for the precise derivation of the analogous result in the original $SU(5)$ Georgi-Glashow model~\cite{Georgi:1974sy}}. To be precise, in this work we will assume the usual picture of symmetry restoration at high temperature above the unification scale (see the discussion in subsection~\ref{whathappens}). 

What happens is that, unless $m_{\tilde\nu_R}$ is large enough, the system will be forever caught in the unbroken phase, even at low temperatures.  
We performed a detailed analysis of radiative corrections in Appendix~\ref{sec:CW} and we found 
\begin{equation}
\label{eq:mtildelowerbound}
m_{\tilde \nu_R} \gtrsim M_{\rm PS}/4\,.
\end{equation}
Combining this with the lower bound on the domain wall tension \eqref{eq:sigmalower} we arrive at
\begin{equation}
\label{eq:sigmalowerfinal}
    \sigma_{\rm Wall} \gtrsim M_{\rm PS}^3\,.
\end{equation}

These topological defects can only be produced at high temperatures, on the order of the unification scale, which takes us naturally to the question about their cosmological history.

\section{What happens in the early Universe?}\label{whathappens}

In order to answer this question, we would have to know the form of the expansion of the Universe 
at astronomically large temperatures on the order of the PS scale - but we don't, and in all honesty, maybe we should stay silent on the issue. The trouble is that we cannot. Thus, we make the plea to the reader to let us assume, for the sake of concreteness, that the standard cosmological model, the big-bang picture, remains valid even at these energy scales. 

\subsection{What is the symmetry at high temperature?}

We know the answer in the SM: the symmetry gets restored. This is what we would expect intuitively - after all, symmetry restoration means less order or more entropy. If the same was true also in the PS theory, we would end up with the so-called Kibble mechanism~\cite{Kibble:1976sj} describing the formation of topological defects at the phase transition from the unbroken to the broken phase.

However, as we said in the introduction, in general symmetries may not be restored at high temperature, at least not at the scale of zero temperature symmetry breaking~\cite{Weinberg:1974hy,Mohapatra:1979qt,Mohapatra:1979vr,Mohapatra:1979bt}. If so, the usual Kibble production of domain walls~\cite{Dvali:1995cc} and  monopoles~\cite{Dvali:1995cj} would not take place, and the fate of topological defects would simply be unknown. If so, one could not claim the existence of domain wall and monopole problems. 

It is a straightforward exercise to show that in this minimal PS theory the original gauge symmetry may not be restored. 
In order to illustrate the situation, it suffices to  consider the simplified potential
\begin{equation}
\label{eq:Vsimplified}
\begin{split}
    V_{\rm simplified} &= V_H + \lambda_{15}\,{\rm Tr}\left[\Phi_{15}^\dagger \Phi_{15}\right]^2 \\
    &\quad \quad -\lambda_{H\Phi} \left(H_{\rm L}^\dagger H_{\rm L} +H_{\rm R}^\dagger H_{\rm R}  \right){\rm Tr}\left[\Phi_{15}^\dagger \Phi_{15}\right]\,.
    \end{split}
\end{equation}
where $V_H$ is given in~\eqref{eq:pot} and we include only $\Phi_{15}$, since it will have the dominant effect over $\Phi_{1}$ due to the much larger number of states. There are more crossed terms mixing $\Phi_{15}$ with $H_{\rm L},H_{\rm R}$ fields, and their presence does not affect our conclusions,
so we ignore them here for simplicity.

 The bottom line is that for the sufficiently large $\lambda_{H\Phi}$,
the original symmetry does not get restored. 
The details of the high-temperature symmetry breaking analysis together with the full expression for $\lambda_{H\Phi}$ bound are given in the Appendix~\ref{sec:thermal}. It tunrs out that the largest coupling, $\lambda_{15}$, needs to satisfy approximately,
\begin{equation}
\begin{split}
  \lambda_{15}\gtrsim \frac{17\pi}{100}(2\alpha_2 + 5 \alpha_4)\,,
  \end{split}
\end{equation}
which at first glance seems well within the perturbative regime. However, due to the large number of real fields, $N=120$ in $\Phi_{15}$, the effective coupling $\lambda_{15}\,N$ becomes of order one, signaling the potential breakdown of perturbation theory. This is similar to the situation encountered in the $SU(5)$ theory with analogous large representation~\cite{Dvali:1995cj}, where a careful analysis shows that this leading order result might not be trusted~\cite{Bimonte:1995xs}. 

For this reason, we shy away from the possibility of symmetry non-restoration at high $T$. There is no such problem in the case of symmetry restoration since it works for any positive value of $\lambda_{H\Phi}$ (or a small negative value), and thus no scalar coupling needs to be large. 

This seems to be a generic pattern of high-temperature thermal potential: in order to non-restore the original symmetry, one needs large representations, which drive effective scalar couplings to be dangerously large. In short, this is yet another example where symmetry restoration, as expected intuitively, seems to be the only option.

\subsection{Symmetry restoration and the fate of topological defects}

Since it appears that the symmetry does get restored, topological defects, both monopoles and domain walls, are then produced via Kibble-Zurek mechanism~\cite{Kibble:1976sj,Zurek:1985qw}. Therefore, due to a cosmological phase transition  in the early Universe, one expects at least an $\mathcal{O}(1)$ of such defects per horizon. 

\subsubsection{Evolution of domain walls} 
The production  of domain walls potentially leads to a cosmological disaster~\cite{Zeldovich:1974uw}. In fact,
 regardless of their initial density, domain walls are expected to enter into a scaling regime, with energy density (see e.g., \cite{Saikawa:2017hiv} for a review) 
\begin{equation}
\label{eq:rhodwevol}
    \rho_{\rm wall}(t)\simeq \sigma_{\rm wall}/t\,,
\end{equation}
As discussed in subsection~\ref{subsubsec:dws}, cosmological consistency of the thermal phase transition
requires $\sigma \gtrsim M_{\rm PS}^3$. 
It follows that such domain wall necessarily dominate the energy of the Universe very fast, leading to unacceptable cosmological scenario. 
It is thus mandatory that domain walls decay-annihilate before that happens.
It is easy to see that this implies the lower bound on the annihilation temperature
\begin{equation}
\label{eq:Tannlower}
    T_{\rm ann} \gtrsim  10^{10}\rm GeV \left(\frac{g_\star (T_{\rm ann})}{10}\right)^{-\frac{1}{4}}\left(\frac{\sigma_{wall} }{\left(10^{13}\rm GeV\right)^3}\right)^{\frac{1}{2}}\,,
\end{equation}
where we normalize the wall tension to its largest possible value compatible with cosmology (see discussion below).

Given the sensibility of $\sigma_{\rm Wall}$ on PS scale~\eqref{eq:sigmalowerfinal} we obtain
\begin{equation}
   T_{\rm ann} \gtrsim 10^{8}\,\rm GeV \,.
\end{equation}
where we used the lower bound on $M_{\rm PS}\gtrsim 4 \times 10^{11}\, \rm GeV$ from~\eqref{eq:mpslowerrunning}.

The trouble is that the domain walls, being topological in nature, are stable - and if one were to take the PS theory at face value, we would be left with the cosmological catastrophe. Since there is no option of inflation at hand, a seemingly desperate measure is called for - violate explicitly the left-right symmetry~\cite{Vilenkin:1984ib}. However, as we discussed above, quantum gravitational effects need not respect global symmetries.  

 For example, a typical leading order higher-dimensional operator responsible for the explicit breaking of parity is schematically given by
\begin{equation}
    \frac{1}{\Lambda^2} (H_{\rm R}^\dagger H_{\rm R})^3\,,
\end{equation}
which amounts an explicit breaking mentioned above $\epsilon = (M_{\rm PS}/\Lambda)^2$, c.f., discussion in subsection~\ref{subsubs:fermionmassandmixing}. The decay of the domain wall is computed by requiring that such operator becomes comparable to the cosmological energy density~\eqref{eq:rhodwevol}, leading to
\begin{equation}\label{eq:cutoffmpl}
    \Lambda \simeq \sqrt \frac{M_{\rm PS}^3 M_{\rm Pl}}{ T_{\rm ann}^2} \lesssim M_{\rm Pl} \,,
\end{equation}
where the upper bound on $T_{\rm ann}$~\eqref{eq:Tannlower}, has been used in the last inequality.

It is reassuring that even such tiny effects suppressed by the Planck scale suffice to do the job~\cite{Rai:1992xw} - and moreover, they do not affect any of the phenomenological predictions of the theory.

\subsubsection{Evolution of monopoles}
In the Kibble picture, causality constrains the correlation length of the scalar field, implying roughly one topological defect (including monopoles) per horizon, which translates into the monopole density over entropy $s$ 
\begin{equation}
    \label{eq:kibble} 
  \frac{n_{\rm Kibble}}{s}\gtrsim\left(\frac{T_{\rm c}}{M_{\rm Pl}}\right)^3,
\end{equation}
where $T_{\rm c}$ is the critical temperature and $M_{\rm Pl}$ is the Planck mass. Zurek later refined this estimate for the case of a second-order phase transition, showing that the relevant correlation length could be even smaller~\cite{Zurek:1985qw} (a result confirmed in various laboratory experiments~\cite{Bowick:1992rz,Dodd:1998aan,Ruutu:1995qz,Baeuerle:1996zz,PhysRevLett.83.5210,Carmi:2000zz,Monaco:2002zz,Maniv:2003zz,Sadler:2006cok}), thus exacerbating the monopole problem. 

Notice that for such high-densities, monopole--antimonopole annihilation in the thermal plasma reduces the density as pointed by Preskill~\cite{Preskill:1979zi}. However, 
additional details of the phase transition (e.g., whether it is first order) can affect the correlation length, potentially pushing it to be of order horizon. Therefore, \eqref{eq:kibble} is a reliable lower bound. If it is saturated, Preskill's correction is not operative given the range of critical temperatures in this PS model. 

The most stringent constraint for PS monopoles follows from direct searches on the flux of monopoles at Earth which is 
\begin{equation}
    \label{eq:flux} 
  {\rm Flux} \simeq n_{\rm M}\, v_{\rm M}  \,.
\end{equation} 
The velocity of monopoles $v_{\rm M}$ in the Galaxy is expected to be greater equal than $10^{-3}$~\cite{Vilenkin:2000jqa}. Monopoles are further accelerated by the magnetic field of the Galaxy. For $10^{11}\rm GeV\lesssim M_{\rm M}\lesssim 10^{17}\rm GeV$, $B\sim 3\,\rm \mu \rm G$, $v_{\rm M}\sim \left(10^{11}\, \rm GeV/M_{\rm M} \right)^{1/2}$~\cite{PDG2024}. Notice that PS monopole mass is roughly $M_{\rm M}\sim 10\, M_{\rm PS}$.

Current constraints on the flux at Earth are given by MACRO collaboration ${\rm Flux}\lesssim 10^{-16}{\rm cm}^{-2}{\rm s}^{-1}{\rm sr}^{-1}$~\cite{MACRO:2002jdv} which is basically independent of $v_{\rm M}$. Furthermore, for relativistic monopoles, ANTARES~\cite{ANTARES:2022zbr} and IceCube~\cite{IceCube:2015agw,IceCube:2021eye} improve the bounds by up to three orders of magnitudes. From the running, we know $M_{\rm M}\gtrsim 10^{12}\,\rm GeV$, implying that in our region of parameter space the monopoles are always, at best, mildly relativistic. 

It therefore suffices to consider here the MACRO bound, which, using \eqref{eq:flux} gives
\begin{equation}
\label{eq:macrobound}
 \text{MACRO bound: \,\,} \frac{n_{\rm M}}{s}\lesssim \frac{10^{-28}}{v_{\rm M}}\,,
\end{equation}
where we plugged in the entropy density of the Universe $s \simeq 400\, {\rm cm}^{-3}$. Considering the range of $M_{\rm PS}$ and how this impacts the velocity $v_{\rm M}$, we conservatively bound $T_{\rm c}$ adopting the density estimate \eqref{eq:kibble} as\footnote{For monopoles that catalyze proton decay~\cite{Callan:1982au,Rubakov:1982fp,Rephaeli:1982nv}, as in theories that violate baryon number, the situation is even more dire, requiring further suppression~\cite{Kolb:1982si,Dimopoulos:1982cz,Freese:1983hz,Kolb:1984yw,Harvey:1982py,Freese:1998es,Arafune:1983tr}. This, however, does not apply to this particular PS theory.}
\begin{equation}
T_{\rm c}\simeq M_{\rm PS}\lesssim 10^{10}
\,\mathrm{GeV}\,.
\end{equation}
Notice that in the model considered here, there isn't sufficient freedom in the unification scale to avoid bound \eqref{eq:macrobound} since $M_{\rm PS}$ is simply too high \eqref{eq:mpslowerrunning}\footnote{Strictly speaking there is a potential loophole when $m_{\tilde\nu_R}$ is fine-tuned nearby inequality \eqref{eq:mtildelowerbound}. In that case, it might be that $T_{\rm c}\ll M_{\rm PS}$ leading to a super-cooled phase transition. We shy away from such a conspiracy. 
 }.

\section{Domain walls sweep monopoles away?}\label{sec:domainmonopolesweeping}

The authors of~\cite{Dvali:1997sa} make a daring suggestion 
that in the decay process, domain walls could actually sweep the monopoles away, which could solve both the domain wall and monopole problems with one shot. If this were to work, it would be a remarkable result, implying that there was never the monopole problem to start with, even with all the assumptions of the standard cosmological model. In particular, there would be no need for inflation from the particle physics point of view. 

\begin{figure*}[th!]
\centering
    \includegraphics[width=1\textwidth]{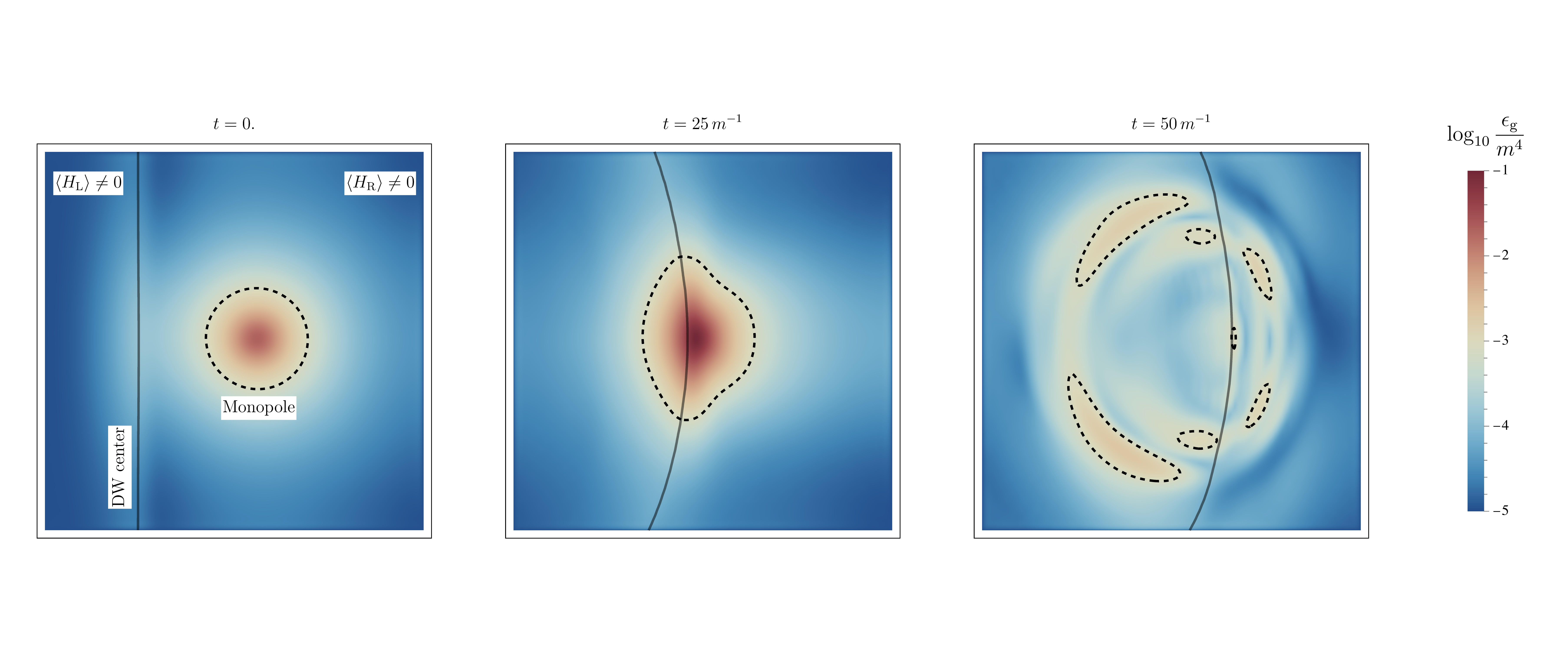}
    \caption{Snapshots of 2D slices of 't Hooft monopole~\cite{tHooft:1975psz} in PS model~\cite{Pati:1974yy} being erased by a left-right domain wall. Colors encode the energy density $\epsilon_{\rm g}$ of the gauge fields, while black dashed iso-line is shown for $\epsilon_{\rm g} =10^{-3}m^4$. The black line represents the center of the domain wall $H_{\rm L}^\dagger H_{\rm L} -H_{\rm R}^\dagger H_{\rm R} =0$. Couplings have been chosen of order one. 
  }
    \label{fig:snapshot}
\end{figure*}

The proposal of~\cite{Dvali:1997sa} was made originally in the context of the minimal $SU(5)$ theory\footnote{The mechanism of monopole sweeping has been subject of several numerical studies for $SU(5)$ gauge symmetric group~\cite{Pogosian:1999zi,Brush:2015vda} and, more recently, for the case of $SU(2)$ symmetry group~\cite{Bachmaier:2023zmq,Valbuena-Bermudez:2023rip}. More in general, its study~\cite{Dvali:2022rgx,2785855} is further motivated by the question of how branes of different dimensionalities interact~\cite{Dvali:2002fi,Dvali:2007nm}. }, but in a sense this mechanism is tailor made for the PS theory, where the discrete LR symmetry is at its core, and it must be dealt with due to the cosmological trouble that its subsequent spontaneous breaking brings upon. On the other hand, in $SU(5)$ there is strictly speaking no discrete symmetry, so one appeals to the portion of the parameter space where there is an approximate $\mathbb{Z}_2$ symmetry - and this is done just to argue that then monopoles could be swept away as the walls decay. It is noteworthy that the huge scale of grand unification, needed for the sake of proton longevity, forces one to deal with temperatures where even a very notion of equilibrium may not hold, due to the fast expansion of the Universe compared to effective interaction rates~\cite{Elmfors:1993pz}.

In any case, the question as to whether the mechanism of sweeping monopoles away is sufficiently efficient in solving the monopole problem, is complicated for a series of reasons. First, the probability $p$ that a monopole encounters a domain wall at least once throughout the cosmological history is hard to compute. Still, it is plausible to expect that this probability is somewhat close to $1$ for several reasons. Firstly,  the domain walls sweeps the Universe with a curvature radius of horizon size. This already brings $p$ close to $1$, especially if domain walls are relativistic, compatibly with a smaller $T_{\rm ann}$. Second, the complicated network dynamics at decay-time can further increase $p$. 

However, depending on the relative internal orientation between the monopole and the domain wall, in some cases no erasure takes place and the monopole simply passes through the domain wall~\cite{Pogosian:2000xv,Pogosian:1999zi}. 
This issue plagues the scenario in its original $SU(5)$ realization, thus requiring a detailed separate study of the cosmological evolution of the domain-wall network which is, however, beyond the scope of the present work.

In order to proceed, one has to tackle first the intricacies of the monopole-domain wall interaction in this PS model which we now turn to.

\subsubsection{Monopole-domain wall interaction }\label{subsec:erasure}

The explicit breaking of parity - which leads to decay of domain walls at $T\simeq T_{\rm ann}$ - leads to an effective pressure that accelerates the walls. A natural question is the fate of the monopole as it encounters one such object.

We provide further details on the nature of the domain wall and how it might affect the interaction of the monopole given their relative orientation in Appendix~\ref{app:numerics}. While these might be rather qualitative arguments, we support them with numerical simulations, of which some details are also reported in Appendix~\ref{app:numerics}.

In Fig.~\ref{fig:snapshot} different frames of 2-D slices of the dynamics are shown. The colors indicate the energy density of the gauge fields, $\epsilon_{\rm g}$, while the black line denotes the center of the domain wall, for which $H_{\rm L}^\dagger H_{\rm L} -H_{\rm R}^\dagger H_{\rm R}  = 0$. Dashed iso-line is also shown for $\epsilon_{\rm g}= 10^{-3}m^4$.
The first panel displays the starting configuration obtained with a numerical relaxation procedure. The domain wall is then accelerated by turning on an explicitly violating parity term at the level of the scalar masses, of the form $H_{\rm L}^\dagger H_{\rm L} -H_{\rm R}^\dagger H_{\rm R}$. 
As it can be seen, upon hitting the wall the monopole immediately unwinds and gauge bosons scatter around (visuals of the dynamics can also be found at the following \href{https://youtu.be/p7-EksvIrAs}{URL}). In the specific simulation shown in the figure, we chose $\lambda'_{\rm LR}>0$. We comment on the case with opposite sign in the Appendix~\ref{app:numerics}.

Another question is wether, in the limit of a highly-energetic scattering, a monopole can be recreated on the opposite side. The scattering of ultra-relativistic monopoles has been tackled in~\cite{Dvali:2022vwh,Zantedeschi:2022czs}, where it was shown that this is unlikely due to the loss of coherence in the annihilation process. Due to this, the entropy suppression makes the recreation of a monopole pair highly improbable.
As expected, the numerical simulation confirms our expectation that the monopole is erased when interacting with the domain wall. This was verified to hold true for different relative orientations between the domain wall and the monopole as well.

\section{Cosmological Consequences} \label{cosmological consequences}


\subsection{How many monopoles left?}

 One might think that the domain wall sweeping might leave the Universe deprived of monopoles. However, at decay time the Universe is in a configuration very similar to the final moments of a phase transition.  
 Namely, upon decaying, the domain walls will engulf regions of false vacuum - corresponding to broken $SU(2)_{\rm L}$ regions - of roughly the size of the horizon. This effectively leads to a correlation length for $H_{\rm R}$ - the remaining non-vanishing order parameter - at most of order horizon size.  
 
The above statement corresponds to the fact that casually disconnected regions will have different orientations in field space, leading once again to the localization of monopoles at the decay time of the domain walls. Therefore, $\mathcal{O}(1)$ monopole per horizon localizes.
This leads to Kibble original density estimate of one monopole per horizon~\cite{Kibble:1976sj,Zurek:1985qw,Preskill:1979zi}, however at a delayed time compared to the monopole production time
\begin{equation}
   \frac{n_{\rm erasure}}{s}\gtrsim \left(\frac{T_{\rm ann}}{M_{\rm pl}} \right)^3\,.
\end{equation}

As discussed in the previous Section, current bounds from monopole direct searches imply $T< 10^{10}\,\rm GeV$, c.f., \eqref{eq:macrobound}.
From the condition imposed by cosmological viability, i.e., domain walls non-domination \eqref{eq:Tannlower}, we obtain 
\begin{equation}
    M_{\rm PS}\simeq \sigma_{\rm Wall}^{1/3}\lesssim 10^{13}\,\rm GeV\,,
\end{equation}
which justifies the chosen normalization in our equations. 
As we will see next, this leads to potentially observable gravitational wave signatures. 

\subsection{Gravitational Waves} \label{sec:GW}

While this scenario is quite natural, one may still wonder about its physical imprints, 
given that the associated scales are far beyond direct experimental reach. Probing 
the thermal history of the Universe is a challenging endeavor; fortunately, decaying 
domain walls can leave a significant footprint in the form of GWs. 
Such signals from domain walls have been studied extensively, and we summarize the 
essential results here. An important feature is that theoretical consistency forces 
the GW energy density into a limited range that may be testable in current and future 
experiments.

The peak of the gravitational-wave spectrum can be estimate following, for example, 
Ref.~\cite{Saikawa:2017hiv} which leads to
\begin{equation}
\label{eq:GWpeak}
\begin{split}
&\left(\Omega_{\rm GW}h^2\right)_{\rm peak} = 3 \times 10^{-7} \, \\&\left(\frac{g_{\star s}(T_{\rm ann})}{100} \right)^{-4/3}
\left(\frac{\sigma_{\rm Wall}}{\left(10^{13}\rm GeV\right)^3} \right)^2 \left(\frac{10^{10}\rm GeV}{T_{\rm ann}} \right)^4\,,
\end{split}
\end{equation}
where $h = H_0 / \bigl(100\,\mathrm{km}\,\mathrm{s}^{-1}\,\mathrm{Mpc}^{-1}\bigr)$ is 
the reduced Hubble constant, 
$g_{\star s}$ is the number of entropic degrees 
of freedom. 

The peak frequency satisfies
\begin{equation}
\label{eq:frequency}
\begin{split}
    &f_{\rm peak }\simeq \left(\frac{R(t_{\rm ann})}{R(t_0)} \right)H(t_{\rm ann})=\\
   & 10^3\,{\rm Hz}\left(\frac{g_{\star}(T_{\rm ann})}{100} \right)^{\frac{1}{2}}\left(\frac{100}{g_{\star s}(T_{\rm ann})}\right)^{\frac{1}{3}}\left(\frac{T_{\rm ann}}{10^{10} \rm GeV}\right)\,,
    \end{split}
\end{equation}
where $R(t)$ denotes the scale factor evaluated at time $t$. 
At lower and higher frequencies than $f_{\rm peak}$, the spectrum is expected to 
fall off as $f^{-3}$ and $f^{-1}$, respectively. 

\begin{figure}[th!]
    \centering
    \includegraphics[width = 1\linewidth]{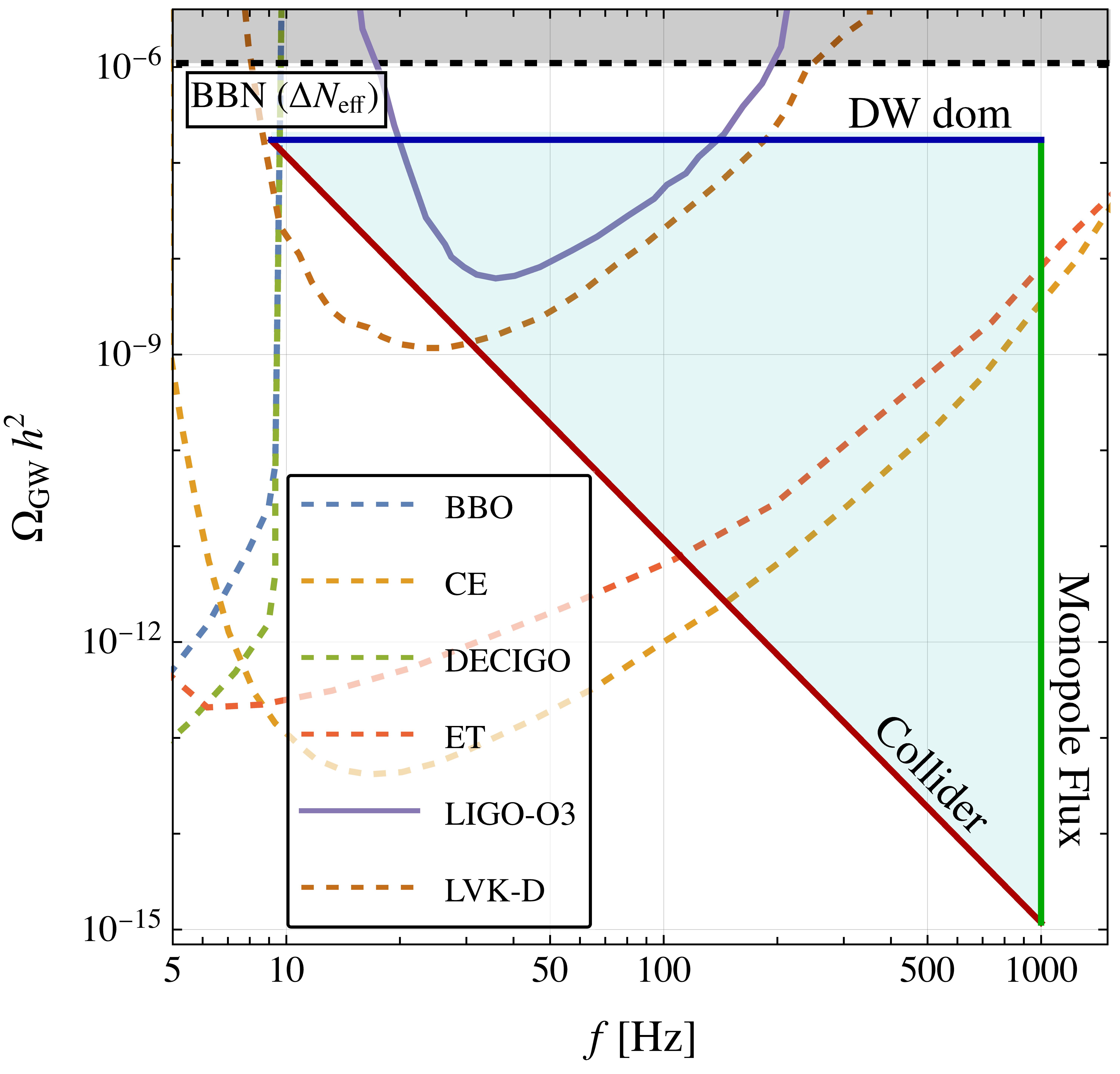}
    \caption{Schematic representation of the gravitational-wave parameter space. 
    The cyan band corresponds to the peak region of the signal, along with the frequency 
    scaling described in the main text. The blue, red, and green lines represent 
    boundaries arising from domain-wall cosmological domination, collider constraints, 
    and monopole direct-search limits, respectively. 
    Note that in the lower portion of the cyan band, the 
    domain walls are non-relativistic and likely cannot efficiently erase monopoles; 
    consequently, we expect the signal to lie very close to the region denoted by 
    the blue line. For details about current experimental reach and future prospects, 
    see the main text.}
    \label{fig:gws}
\end{figure}

From the theoretical arguments presented earlier, it follows that
\begin{equation}
\begin{split}
   &4\times 10^{11}\, {\rm GeV}\lesssim \sigma_{\rm Wall}^{1/3}\sim M_{\rm PS}\lesssim 10^{13}\,\rm GeV\,,\\
   \,\\
   & \qquad \qquad 10^{8}\, {\rm GeV}\lesssim T_{\rm ann}\lesssim 10^{10}\,\rm GeV\,.
   \end{split}
\end{equation}
Because $T_{\rm ann}$ sets the frequency scale in Eq.~\eqref{eq:frequency}, 
we see that the GW peak must lie between about 
\begin{equation}
    10\,{\rm Hz} \lesssim f_{\rm peak}\lesssim 10^{3}\,\mathrm{Hz}\,.
\end{equation}
This range is highlighted in Fig.~\ref{fig:gws}.

A reasonable requirement for efficient monopole erasure is that the domain walls become 
relativistic, implying 
\begin{equation}\label{eq:tanntom}
T_{\rm ann}\simeq (\sigma_{\rm Wall}/M_{\rm Pl})^{1/2}\equiv T_{\rm dom}\,. 
\end{equation}
This choice - corresponding to  explicit LR symmetry violation, with a cutoff of order $M_{\rm Pl}$, as seen in \eqref{eq:cutoffmpl} -  pushes the Universe to the verge of a cosmological catastrophe, as the 
domain walls almost dominate its energy density. Under these conditions, the amplitude of the signal becomes  independent of the underlying model parameters and is represented by the 
horizontal blue line in Fig.~\ref{fig:gws}. The entire possible range of peak values 
is shown by the cyan band. The continuous red line corresponds to the lowest viable 
$M_{\rm PS}$ (as suggested by the running analysis) in correspondence of which a new scalar states with SM numbers $(3_{\rm C},2_{\rm L})_{7/6\rm Y}$ lies around  
$10\,$TeV. Potential friction effects along the lines of the discussion of Ref.~\cite{Nakayama:2016gxi} would extend the lowest value of $\Omega_{\rm GW}$ by roughly four orders of magnitude. We shy away from a complete characterization of this regime since  moving away from the blue line (and therefore from \eqref{eq:tanntom}) likely reintroduces the monopole problem. 

For completeness, we further report existing bounds and future prospects. 
The dashed black line shows the bound from CMB, which stems from the fact that too much energy density in gravitational waves would constitute radiation contributing to $\Delta N_{\rm eff}$ (see e.g., \cite{Aggarwal:2025noe}). The colored dashed lines indicate the reach of future-generation gravitational wave experiments such as Big Bang Observer~\cite{Harry:2006fi}, Cosmic Explorer~\cite{Evans:2021gyd}, DECIGO~\cite{Kawamura:2020pcg} and Einstein Telescope~\cite{Punturo:2010zz}. Finally, the continuous line indicates the present reach of LIGO-O3 and the reach right below corresponds to the prospect of LVK-D~\cite{aLIGO:2020wna,VIRGO:2014yos,KAGRA:2018plz}.

\subsection{Dark Matter}

Since the PS theory has all the attributes of a self-contained theory, one must ask whether it provides a possible candidate for the dark matter of the Universe. The answer is affirmative, since the theory offers a natural WIMP contestant in a popular form of the inert SM scalar doublet - the scalar analog of the lepton doublet field, $\tilde \ell_L$. Since it is not coupled to the fermions, it can be sufficiently long lived to be the dark matter today, as long as it does not have direct linear couplings with other scalars. To be more precise, the couplings that must be small are of the type $H_{\rm L} \Phi H_{\rm R}$, which ensures the relative stability of the sneutrino (scalar analogue of neutrino) inside $H_L$, and the smallness of its tadpole induced vev.

In the limit when its direct linear coupling go to zero, one has an accidental protective $\mathbb{Z}_2$ symmetry, which guarantees its stability - or its longevity, if not exact. 
The end result is that the masses of the inert doublet states must lie below roughly TeV, see e.g.,~\cite{LopezHonorez:2006gr,Dolle:2009fn}. In other words, if not already at the LHC, it would be accessible at the next hadron collider. 

\subsection{Genesis}

Since the authors do not adhere to any of the religious paradigms of genesis (they are open minded though), they are compelled to look for alternative explanations. There are two scenarios that have dominated the field of genesis in particle physics over the years, the leptogenesis~\cite{Fukugita:1986hr} and the electroweak baryogenesis~\cite{Kuzmin:1985mm}. The former one is absent here - we are in the neutrino Dirac picture with conserved baryon and lepton numbers. The only other alternative is the electroweak genesis, and it is surely possible. As we said, the unification conditions leave the masses of the plethora of SM Higgs doublets free, and all that is needed then is an additional light doublet - which can then account for the first order phase transition at the electroweak temperatures. 

We have seen that such a doublet, analogue of a lepton doublet, makes a natural dark matter candidate. It would be interesting if it was the only light doublet, besides the SM Higgs one, and thus also played the role of establishing the first order phase transition~\cite{Chowdhury:2011ga}. This would naturally fix the masses of this inert doublet, with the dark matter candidate mass between $45-80\,$GeV, and the other states in the doublet below $350\,$GeV~\cite{Chowdhury:2011ga} - potentially accessible at the LHC, and surely at the next hadron collider. It is worth commenting that this model predicted the SM Higgs boson mass to lie below $130\,$GeV~\cite{Chowdhury:2011ga}.
Of course, since there are many doublets available, no prediction can be made as to which one does the job. 

A comment is noteworthy. In a true theory of genesis, the baryon and lepton numbers of the Universe should not depend on the initial conditions. This is true, at least in principle, in the leptogenesis scenario. Namely, at sufficiently high temperatures, the heavy Majorana RH neutrinos can be in equilibrium, and their interactions can wash out the original $B-L$ charge of the Universe. Similarly, sphaleron interactions at high temperature wash out the original $B+L$ number of the universe, and so once they get out of equilibrium, the RH neutrinos can produce the net lepton number of the universe, which through sphaleron interactions gets transmitted also to the baryon number. 

 On the other hand, electroweak baryogenesis can compute only the $B+L$ combination, and so it does not provide a complete scenario, since $B-L$ remains an initial condition. It does leave the imprint of gravitational waves   
in the expected frequency range $10^{-5}\text{--}1\,\mathrm{Hz}$~\cite{Dorsch:2016nrg,Wang:2019pet,Zhou:2020irf},
 potentially within reach of 
LISA~\cite{Caprini:2015zlo,LISA:2017pwj,Caprini:2019egz} 
and DECIGO~\cite{Kawamura:2020pcg}. This signal would therefore be distinct from the 
one associated with domain walls.



\section{Summary and outlook}\label{sec:concluion}

  In this work, we performed a detailed study of a minimal renormalisable PS theory with the unification Higgs sector that leads to the neutrino Dirac picture. The main features of the theory are the simplicity and completeness - the theory can account for all the phenomena beyond the SM physics. It possesses a natural dark matter candidate in the form of the inert doublet and allows for the electroweak baryogenesis. 

  It turns out that the unification scale is huge, above approximately $10^{11}\,$GeV, hopelessly out of direct experimental reach. Fortunately, the theory leaves the imprints in the form of possibly light scalar states, the analogue of the down quark singlet, besides the number of weak doublets - these states drop out from the unification constraints. 

The crucial role in all of this is played by the interplay of magnetic monopoles and domain walls, that arise and interact during the high-scale cosmological phase transition. These monopoles  carry fractional electric charge due to the unification of color and $B-L$ in $SU(4)_{\rm C}$. Due to the huge size of the unification scale, current experimental bounds on their allowed flux would be violated. 

Fortunately, the spontaneous breaking of the original  discrete left-right symmetry further produces domain walls that must decay and in the process can ``sweep'' these monopoles away. We have carefully addressed the dynamics of monopoles encountering a domain wall, and showed that they unwind. The reason for this is analogous to the original suggestion in the context of the minimal $SU(5)$ theory~\cite{Dvali:1995cj}. 

The requirement that domain walls do not dominate the Universe constraints their annihilation temperature, leading to the production of gravitational waves in the range of frequencies between roughly $10-10^3\,\rm Hz$. The upper bound stems from the above-mentioned constraint on the monopole flux, while the lower one would imply the presence of a scalar colored weak doublet, with SM quantum numbers $(3_{\rm C},2_{\rm L})_{{\pm 7/6}_{\rm Y}}$, lying below $10\,\rm TeV$ energies. 

We have focused here on the smallest possible unification Higgs sector - fundamental representations of both $SU(4)_{\rm C}$ and $SU(2)_{\rm L,R}$ - that took us to the Dirac neutrino picture. A natural question is what happens in the different case of Majorana neutrinos, which requires an alternative Higgs scalar sector. In particular, 
in the renormalizable version of the theory, a suitable modification involves swapping the fundamental representations for the symmetric ones, which allows the unification scale to be as low as $10^{9}\mathrm{GeV}$~\cite{Preda:2025afo}. Notably, below $10^{10}\mathrm{GeV}$, constraints on monopole flux from the MACRO experiment are avoided — at least according to Kibble's estimate of $\sim \mathcal{O}(1)$ monopole per horizon at the production time. 

Therefore, within this parameter range, monopole erasure is not strictly necessary and domain walls could potentially decay immediately after formation, significantly suppressing the GW signal. However, such a low scale would imply the presence of new light scalar at nearby energies~\cite{Preda:2025afo}. For unification scales above $10^{10}\mathrm{GeV}$, we expect the results presented here to hold similarly. It is interesting to note that in this realization of PS symmetry, the GW frequency peak might fall below the Hz range, enhancing experimental detectability as illustrated in Fig.~\ref{fig:gws}.

In contrast, the non-renormalizable version of the theory, discussed in Ref.~\cite{Preda:2025afo}, does not require additional scalar degrees of freedom. However, neutrino mass consistency pushes the unification scale significantly higher, to around $10^{13}\mathrm{GeV}$. At this higher scale, cosmological considerations place the model on the edge: the resulting gravitational wave signal is fixed around frequencies of $10^{3}\mathrm{Hz}$ and its amplitude is necessarily maximized, compatibly with domain wall near energy-domination of the Universe. Moreover, this scenario implies a Universe currently populated with monopoles just below observational limits. These intriguing aspects of the Majorana framework remain open for future exploration.

Another interesting question is what happens in the grand unified extension of the PS theory. 
There the picture gets even more involved, with a rich dynamics of topological defects, that includes further cosmic strings and even domain walls bounded by strings \cite{Kibble:1982ae,Kibble:1982dd}. The interplay between these toplogical objects and gravitational waves has been recently addressed e.g., in~\cite{Maji:2024tzg,Maji:2025yms,Lazarides:2023iim}.

Coming back to PS model, one may wonder about the implications if a right-handed charged gauge boson $W_{\rm R}$ were discovered at the next-generation collider. Such a discovery would invalidate the current theoretical framework, 
raising the question of how quark-lepton unification could still be preserved. The answer lies in employing the scalar field $\Sigma = (15_{\rm C},1_{\rm L},1_{\rm R})$ to break the original PS symmetry down to the LR one. The cosmological consequences depend critically on whether $\Sigma$ is parity-even or parity-odd. If $\Sigma$ was parity-even, domain walls would be produced at the LR breaking scale; if parity-odd, their production would occur at $M_{\rm PS}$, as discussed previously. In the parity-even scenario, an important question arises: could the decay of these domain walls, necessary for reducing monopole abundance, potentially explain the stochastic gravitational wave background recently detected by pulsar timing arrays~\cite{NANOGrav:2023gor,EPTA:2023fyk,Reardon:2023gzh,Xu:2023wog} at around $\rm n Hz$ frequencies? These and other critical questions are actively under investigation.

\begin{acknowledgments}
We thank Borut Bajc for useful discussion on the potential perturbativity issues of symmetry non restoration. MZ further acknowledges warm hospitality at the Jožef Stefan institute during the completion of this work. He further thanks Maximillian Bachmaier, Juan Sebastian Valbuena-Bermúdez and Gia Dvali for illuminating discussions on topological defects and their dynamics over the years.
G.S. is grateful to the friendly staff of the Briig hotel cafe in Split for their friendly hospitality during the course of this work. 
\end{acknowledgments}


\appendix


\section{2-loop running}\label{sec:2loop}
Since we are working at two-loop order, the SM gauge couplings $\alpha_{i}$ ($i=1,2,3$) evaluated at the unification scale, are properly mapped to the unification coupling as~\cite{Hall:1980kf,Weinberg:1980wa}
\begin{equation}
\label{lambdaigeneral}
\begin{split}
   & \frac{1}{\alpha_{i}(\mu)} \;=\; \frac{1}{\alpha_{\rm G}(\mu)} \;-\; \tilde\lambda_i(\mu)\,,\\
   & \tilde\lambda_{i}(\mu) \;=\; \frac{1}{12\pi}\left[\,{\rm Tr}\,\bigl(t_{\rm V}^2\bigr)
   \;-\; 21\,{\rm Tr}\,\bigl(t_{\rm V}^2\bigr)\,\log\!\Bigl(\frac{M_{\rm V}}{\mu}\Bigr)\right],
\end{split}
\end{equation}
where the functions $\tilde\lambda_i(\mu)$ arise from the effective gauge theory obtained after integrating out the heavy states, $t_{\rm V}$ are the generators in the representations associated with the heavy gauge bosons, and $M_{\rm V}$ denotes their mass thresholds. Contributions from the scalar sector are neglected because they are numerically suppressed. Moreover there are no additional fermionic contributions at the GUT scale.

In many scenarios, one chooses $\mu = M_{\rm V}$ so that the scale-dependent term in Eq.~\eqref{lambdaigeneral} vanishes. However, in our setup the gauge bosons have different masses, cf.~\eqref{eq:gaugeboson}, so such a straightforward choice is not possible. For simplicity, we chose $M_{\rm PS} = \mu = M_{X_{\rm PS}}$. This leads to
\begin{equation}
    \begin{split}
        &\tilde\lambda_3(M_{\rm PS}) = \frac{1}{12\,\pi}\,,\\
        &\tilde\lambda_{1}(M_{\rm PS}) = \frac{3}{60\,\pi}\left(\frac{19}{6} - \frac{21}{2} \log \frac{M_{W_{\rm R}}}{M_{\rm PS}} \right)\,,
    \end{split}
\end{equation}
which must be properly be included in the PS unification condition \eqref{eq:psunificationcondition}, following \eqref{lambdaigeneral}.

We performed the most general two loop running including proper matching conditions. It turns out that the $M_{\rm PS}= M_{X_{\rm PS}}$ scale we found in the 1-loop analysis is lowered, at most, by roughly $10\%$. Namely, assuming all the new scalar states are at $M_{\rm PS}$, we find $M_{\rm PS}^{2-\rm loop}\simeq 4.6 \times 10^{13}\,\rm GeV$ as opposed to the $M_{\rm PS}^{1-\rm loop}\simeq 5 \times 10^{13}\,\rm GeV$. 

For arbitrary scalar mass thresholds, a light $(3_{\rm C},2_{\rm L})_{7/6_{\rm Y}}$ has the largest effect in lowering $M_{\rm PS}$, just like in the one-loop. Moreover, both the Higgs doublets and the scalar particle with down quantum number do not drop out from the unification condition due to their non trivial two loop coefficients. They have anyway, a marginally small impact on the running. In particular, if the former ones are also light, around $\rm TeV$ scale, we get the lower bound on PS scale
\begin{equation}
    M_{\rm PS}^{2-\rm loop}\gtrsim 4.1\times 10^{11}\rm GeV\,.
\end{equation}
Notice that this is consistent with less than a $10\%$ variation w.r.t. the 1-loop result.

\section{Minimal tension from radiative corrections}\label{sec:CW}
In this section we will show that there is a lower bound on the minimal tension of the domain wall. Naively, one might think that lowering arbitrarily the singlet mass $m_{\tilde\nu}$ does the job. While this is true, viability of the cosmological scenario imposes a lower bound on the mass of the singlet, therefore forcing the tension to be near the GUT scale value. The reason for this is that for too small singlet mass, the radiative corrections to the potential~\cite{Coleman:1973jx} stemming from the gauge sector can not only make the unbroken phase a local minimum, but a global one. Obviously, such situation ought to be avoided cosmologically. 

To estimate for which value of singlet mass this happens, we consider the one-loop effective potential in the limit of vanishing scalar couplings
\begin{equation}
V_{\mathrm{eff}}(\phi) \;=\; V_{\mathrm{tree}}(\phi)
\;+\;
V_{\mathrm{1\text{-}loop}}(\phi),
\end{equation}
where the tree-level potential \(V_{\mathrm{tree}}(\phi)\) is the classical scalar potential,
and the one-loop correction is given by
\begin{equation}
V_{\mathrm{1\text{-}loop}}(\phi)
\;\simeq\;
\frac{1}{64\pi^2}\; \Biggl[
  3 \sum_{\text{vectors}} M_{\rm V}^{4}(\phi)\,\biggl(
      \ln\!\bigl[\tfrac{M_{\rm V}^{2}(\phi)}{\mu^2}\bigr]
      \;-\;\tfrac{5}{6}
    \biggr)
\Biggr]\,,
\end{equation}
where we neglected the scalar contribution and $\mu_r$ is the reference scale. 

We evaluate the gauge boson masses along the vev direction $\langle \Phi_{\rm R\,i}\rangle = \phi = \delta_{i1}\,v_{\rm PS}$. Basically, it suffices to sum over the gauge boson masses \eqref{eq:gaugeboson} - properly counting them - replacing $v$ with $\phi$.

Armed with this, we evaluate the effective potential in the tree level vacuum using $\phi^2 = m^2/(\lambda + \lambda')$ which gives
\begin{equation}
\begin{split}
V_{\mathrm{eff}}(\phi) &= V + V_{1-\rm loop}(\phi)=
\;
\frac{m^{4}}{512\,\pi^{2}\,\bigl(\lambda+\lambda'\bigr)^{2}}\cdot \\
&\;\Bigl[
  -45(2 \alpha_2 + 5\alpha_4)^2 
  - 128(\lambda +\lambda') \\ 
  & + 54(2\alpha_2 + 5 \alpha_4)^2\log \left(\frac{3\pi (2 \alpha_2 + 5 \alpha_4)\mu^2}{2\mu^2 (\lambda + \lambda')} \right)
\Bigr].
\end{split}
\end{equation}

Following~\cite{Linde:1978px}, we require that the energy of the potential at the origin is $V_{\mathrm{eff}}(\phi)\lesssim 0$. This condition ensures that the unbroken phase does not become a global minimum. After fixing the reference scale $\mu$ such that the logarithm is $\mathcal{O}(1)$, we arrive at
\begin{equation}
\label{eq:mnusqbound}
        m_{\tilde \nu_{\rm R}}^2 \gtrsim \frac{9M_{\rm PS}^2}{64\,\pi} \frac{(2 \alpha_2 + 5 \alpha_4)^2}{\alpha_4}\simeq \frac{1}{20} M_{\rm PS}^2\,,
\end{equation}
where in the last equality we considered typical value for the gauge coupling around PS scale, $\alpha_{4}\sim 37^{-1}$ and $\alpha_2 \sim 42^{-1}$. 

Combining bound \eqref{eq:mnusqbound} with the lower bound on the domain wall tension \eqref{eq:sigmalower} we arrive at
\begin{equation}
    \sigma_{\rm Wall} \gtrsim M_{\rm PS}^3\,.
\end{equation} 

\section{Thermal potential}\label{sec:thermal}
Let us consider the simplified potential
\begin{equation}
\begin{split}
    V_{\rm simplified} &= V_H + \lambda_{15}{\rm Tr}\left[\Phi_{15}^\dagger \Phi_{15}\right]^2 \\
    &\quad \quad -\lambda_{H\Phi} \left(H_{\rm L}^\dagger H_{\rm L} +H_{\rm R}^\dagger H_{\rm R}  \right){\rm Tr}\left[\Phi_{15}^\dagger \Phi_{15}\right]\,.
    \end{split}
\end{equation}

For weak couplings, in the perturbative regime, we can compute the effective thermal corrections at $T\gg M_{\rm PS}$ using the condition given by Weinberg~\cite{Weinberg:1974hy}
\begin{equation}
\Delta V(T) = \frac{T^2}{24}\left[\left( \frac{\partial^2V}{\partial \varphi_i\partial\varphi_i}\right) + 3 g^2(T_a T_a)_{ij}\varphi^i\varphi^j \right]\,,
\end{equation}
where $\varphi_i$ correspond to the real components of the fields. The second term is, in fact, generated by the gauge bosons.  

At high temperatures the masses become
\begin{equation}
    \begin{split}
        &m_{H}^2(T) = \frac{T^2}{24}\left[ 36\lambda +24 \lambda' + 16\lambda_{\rm LR} +4\lambda_{\rm LR}'  - 120\lambda_{H\Phi}\right. \\
        &\qquad\qquad \qquad\qquad\qquad \qquad\qquad\left.+ 9\pi \left(2\alpha_2 + 5 \alpha_4\right)\right]\,,\\
        &m_{\Phi_{15}}^2(T) = \frac{T^2}{24}\left[ 244\lambda_{15} - 32 \lambda_{H\Phi} + 24\pi \left(3\alpha_2 + 4 \alpha_4\right)\right]\,.
    \end{split}
\end{equation}

We therefore need to fulfill the condition
\begin{equation}
    \lambda_{H\Phi}\gtrsim \frac{1}{120}[36 \lambda + 24 \lambda' + 16 \lambda_{\rm LR} + 4\lambda_{\rm LR}' + 9\pi(2\alpha_2 + 5 \alpha_4)]\,,
\end{equation}

Boundedness of the potential gives 
\begin{equation}
\label{eq:inequality1}
    \lambda + \lambda'>0\,, \quad \lambda_{15}>0\,,\quad \lambda_{H\Phi}^2 < 4\, \lambda_{15}(\lambda + \lambda')\,.
\end{equation}
Notice that at the same time we need to preserve positivity of masses in vacuum \eqref{eq:masses}. To non restore $H_{\rm R}$ we can work in the limit of vanishingly small - but negative - $\lambda'$ and $\lambda_{\rm LR}'$. In this limit $\lambda_{\rm LR} \gtrsim 2 \lambda$. 

It is then straightforward from \eqref{eq:inequality1}, to obtain
\begin{equation}
    \lambda_{15} > \frac{\left[68 \lambda+ 9\pi (2 \alpha_2 + 5 \alpha_4)  \right]^2}{14 400 \lambda}\,.
\end{equation}
Extremizing over $\lambda$ we find the minimum of the above inequality, which is given by $\lambda =9\pi (2 \alpha_2 + 5\alpha_4)/68$, which leads to
\begin{equation}
\label{eq:l15ineq}
    \lambda_{15}> \frac{17 \pi}{100}(2 \alpha_2 + 5 \alpha_4)\,.
\end{equation}
This is the inequality that we discuss in the main text. Notice that indeed we made approximations and dropped contributions due to scalar couplings $\lambda'$ and $\lambda'_{\rm LR}$. Even in the best-case scenario in which these coupling are negative - but still compatible with the scalar spectrum \eqref{eq:masses} - inequality \eqref{eq:l15ineq} is modified at best by an $\mathcal{O}(1)$ factor, leaving our conclusions unaffected.

\section{More on the domain-wall/monopole interaction}\label{app:numerics}

To support our findings quantitatively we performed numerical simulation of the sweeping of a monopole by a domain wall in the PS model. The initial conditions for both topological configurations are obtained considering the scalar fields ansatz \eqref{eq:monopolethooft} and \eqref{eq:dwansatz} for the monopole and domain wall respectively and then performing a relaxation procedure in fictitious time with a damping term. This is enough to relax the profiles of the scalar and gauge fields. The latter is also forced to align with the scalar field in order to minimize energy. Sommerfield absorbing boundary conditions as well as Lorentz gauge (we performed our simulations also adopting temporal gauge and found no differences) are adopted. The specific values of the scalar couplings are irrelevant in our discussion. Sufficient to say that they were all non-vanishing and chosen to give masses of order $1$ in $m$ units.

To grasp a feeling of the interaction between the domain wall and the monopole, we analyzed the spatially-dependent curvature of the gauge degrees of freedom along the domain wall solution \eqref{eq:dwansatz}. Obviously, the mass of the LR gauge bosons simply swap as we traverse through the wall. This leads to an effective attractive interaction between the domain wall and the monopole which favors the unwinding.  

What about the colored sector?
As discussed in the main text, the sign of the coupling $\lambda'_{\rm LR}$ determines wether $H_{\rm L,R}$ are aligned or perpendicular in color space along the support of the domain wall. This implies either that the symmetry is further broken to $SU(2)_{\rm C}\times U(1)\times U(1)$ for positive $\lambda'_{\rm LR}$ or to $SU(3)_{\rm C}\times U(1)$ in the opposite case. We verified that the scalar sector actually matches the number of would-be Goldstones eaten by the gauge fields. 
 
Depending on the relative orientation of the wall with respect to the monopole (perpendicular or parallel in $SU(4)_{\rm C}$ space), two things can happen: either the colored gauge degrees of freedom composing the monopole become massless as they go through the domain wall analogously to the LR gauge bosons, or they have the same asymptotic masses in both the left and right preserving vacuum, with a small potential dip localizing in the domain wall center. 
In the former case - corresponding to positive $\lambda'_{\rm LR}$ and showed in Fig. \ref{fig:snapshot} - the monopole observes an even stronger attractive interaction wrt the domain wall, and it untwists immediately upon reaching the wall. In the latter case of negative $\lambda'_{\rm LR}$, the mass dip acts as a potential well for the colored gauge bosons which remains partially trapped inside the wall. When numerically simulating this case we can clearly observe such behavior. Of course, nothing prevents the trapped color gauge bosons to leak outside of the domain wall and slowly dissipate. Regardless of this, we observe the untwisting of the monopole upon scattering with the domain wall.

\setlength{\bibsep}{4pt}
\bibliography{citations.bib}
\end{document}